# Focused-ion-beam-induced deposition of superconducting nanowires

E. S. Sadki, S. Ooi, and K. Hirata

National Institute for Materials Science, 1-2-1 Sengen,

Tsukuba, Ibaraki 305-0047, JAPAN

Superconducting nanowires, with a critical temperature of 5.2 K, have been synthesized using an ion-beam-induced deposition, with a Gallium focused ion beam and Tungsten Carboxyl, $W(CO)_6$, as precursor. The films are amorphous, with atomic concentrations of about 40, 40, and 20 % for W, C, and Ga, respectively. Zero Kelvin values of the upper critical field and coherence length of 9.5 T and 5.9 nm, respectively, are deduced from the resistivity data at different applied magnetic fields. The critical current density is $J_c = 1.5 \times 10^5$ A/cm$^2$ at 3 K. This technique can be used as a template-free fabrication method for superconducting devices.

One of the most common applications of focused ion beam (FIB) is the deposition of metallic or insulating films from the induced decomposition of a chemical precursor over a substrate by the ion beam.[1-3] This technique, so-called ion beam induced deposition (IBID), is successfully used to repair electronic devices and photomasks.[4,5] The main advantages of IBID are the deposition of the desired patterns of films without the need of a template, such as a mask or resist, and the ability to correct or modify only parts of the overall circuit design, without restarting the whole lithographic process. Furthermore, one recent promising application of IBID is the growth of three-dimensional (3D) nanostructures.[6] By using appropriate precursors, deposition by IBID have been achieved for a variety of materials, including W, C, Pt, $SiO_x$, Au, Al, and Cu.[7-13] From the above list of materials, it is clear that IBID research is principally driven by the needs of the semiconductor industry, such as fixing integrated circuits (IC) on Silicon chips. However, to our best knowledge, the IBID technique has never been used for the deposition of superconducting materials. This might be due to the absence or difficulty in the preparation of suitable chemical precursors, or the lack of interest from the manufacturers of FIB facilities.

In this letter, we report on the appearance of superconductivity in thin films prepared by the IBID of Tungsten Carboxyl, $W(CO)_6$, and Gallium FIB.[14] The IBID deposited films from $W(CO)_6$ and Ga-source FIB are commonly available options in commercial FIB facilities, and therefore their physical and structural characteristics have been extensively studied in the past.[15-23] However, their properties at low temperatures have not been reported before this work. In these deposition conditions, the nominal film composition is expected to be mainly Tungsten metal, which has a superconducting transition temperature of 0.01 K for bulk single crystals.[24] However, from previous results[15,17,19,22] and our analysis, the deposited films are not crystalline, but are amorphous and always include fractions of Carbon and Gallium elements. Although the origin of superconductivity in these films is not exactly known, we propose that their amorphousity, and the presence of C and Ga play a crucial role in this phenomenon, as it will be discussed later.

For transport measurements, a nanowire has been IBID deposited in a four-point configuration. Temperature dependence of the resistivity of the nanowire has been measured at different applied magnetic fields, from which the upper critical field and coherence length have been deduced. The critical current density was derived from its voltage-current characteristics. A larger film has been synthesized for magnetic measurements and structural analysis by X-ray diffraction (XRD) and electron probe micro analysis (EPMA).

The IBID experiments are carried out using a Micrion 2500 FIB machine, equipped with a $W(CO)_6$ precursor reservoir connected to the FIB vacuum chamber by means of a nozzle. The operating $Ga^+$ ion beam energy is 30 KeV. The deposition FIB current is set to 98 pA, with an aperture size of 100 μm. The precursor temperature is set to 61 °C at a slow rate of 0.5 °C/min to achieve good equilibrium conditions. During the deposition, the chamber pressure is kept at 2.6 $10^{-5}$ Torr. In all the experiments, an FIB dose of 1 nC/μm$^2$ is used, corresponding, from calibration, to a deposited film thickness of 120 nm. Transport and magnetic measurements are performed in a liquid helium cryostat with a superconducting magnet, using a standard AC lock-in amplifier technique, and a Quantum Design MPMS system, respectively.

Figure 1 shows a scanning electron microscopy (SEM) of an IBID deposited nanowire on a Silicon Oxide substrate, connected in a four-point contact configuration to IBID deposited electrodes as well. The wire dimensions are 10 µm in length, and 300, 120 nm, for width and thickness, respectively. Different type of substrates, such as MgO and Sapphire, were also used for the deposition. The resistivity at room temperature is $\rho_n$= 200 µΩ cm, which is consistent with the reported values for IBID deposited films from the same precursor.[22] At low temperatures, the nanowire resistivity at zero field exhibits a superconducting transition at $T_c$= 5.2 K (Fig. 2). The width of the transition is $\Delta T$= 0.5 K. This behavior is found to be independent of the substrate material. Furthermore, the magnetization of a deposited film on MgO (film2), with dimensions of 200 by 200 µm, and 120 nm in thickness, also shows the transition at the same $T_c$. To obtain the upper critical field and coherence length, the resistivity is measured at different applied fields from 1 to 5 T (Fig.2). The resistivity shows pronounced broadening with increasing field as seen in other superconducting thin films.[25] The upper critical field, $B_{c2}$, is defined at 90 % of the transition ($\rho/\rho_n$=0.9). Figure 3 displays a graph of the extracted $B_{c2}$ values as a function of temperature. Due to the small curvature near $T_c$, it was difficult to fit it to the linear $B_{c2}$ equation from Ginzburg-Landau (GL) theory[26] or to the Werthamer formula.[27] However, a good fit of the data to an empirical power law could be achieved, with $B_{c2}(0)$ = 9.5 T. From the GL formula $B_{c2}(0)=\Phi_0/2\pi\xi(0)^2$, the coherence length is $\xi(0)$ = 5.9 nm. The critical current density at 3K, is deduced from the current-voltage characteristic at the sudden jump of the voltage, corresponding to a value of $J_c$=1.5 $10^5$ A/cm$^2$. This is a moderate value comparing to low vortex pinning NbGe ($J_c\sim 10^2$ A/cm$^2$)[28] and strong pinning MgB$_2$ ($J_c\sim 10^7$ A/cm$^2$).[29]

Structural analysis of the films was conducted by XRD and EPMA. The XRD pattern did not show any Bragg peaks confirming the amorphous nature of the films. The results of EPMA study on film2 are summarized in Table I. Mapping on the whole film area, i.e. 200 µm by 200 µm, showed basically a uniform distribution of the elements W, C, and Ga. For comparison, the atomic composition at the center and edge of the film is presented. The uniformity is clear, with average atomic concentrations of about 40, 40, 20 %, for W, C, and Ga, respectively. The concentration of Ga in our films, which is implanted from the ion beam, is similar to the published values in the literature.[17,19,22] However, our data shows a higher C concentration in comparison with 10 to 20 % from references 17, 19, and 22. This excess might be due mainly to the sample surface, since the previously reported chemical analysis by Auger electron spectroscopy (AES),[17,19,22] secondary ion mass spectroscopy (SIMS),[22] and X-ray photoelectron spectroscopy (XPS),[19] have been performed after etching the top surface of the films. Further studies on the structure of the films, and the effect of the deposition conditions are in progress.

Superconductivity in amorphous tungsten films,[30-33] and in its alloys, such as with C,[34,35] Si,[33] and Ge,[33,36] have been reported in the past. First, we rule out that the origin of superconductivity in our films is due to an alloy or compound of W with Si, because the same behavior, i.e. same $T_c$ and $\rho_n$, is reproduced on other substrates than SiO$_2$/Si (for example on MgO and Al$_2$O$_3$). Films of $\beta$-W (A15 structure) deposited by electron beam evaporation on glass or sapphire substrates, with similar thickness to our samples, exhibits superconducting properties, with $T_c$ from 3.2 to 3.35 K.[30,31] Collver and Hammond[32] have showed that by depositing W-films at 4.2 K and warming them up to 300 K, the $T_c$ dropped dramatically from 3.5 K towards the bulk value. This result demonstrates that the amorphousity is a necessary condition for the occurrence of high $T_c$ in W-films. However, the above $T_c$s are still lower than the reported value in our samples. This implies that the elements C and/or Ga contribution is important. Tungsten and Carbon compounds are a hexagonal WC, a high temperature phase fcc WC, and a hexagonal $W_2C$. Hexagonal WC is not superconducting,[34] fcc WC, which is stabilized at room temperature by rapid quenching, and $W_2C$, have superconducting $T_c$ of 10 K and 2.7 K, respectively.[34] The fcc WC phase is not present in our films since it is unstable at room temperature. Amorphous thin films of W and C deposited by RF sputtering, showed superconducting behavior with $T_c \sim 5$ K.[30] The atomic concentrations were 92-94 and 6-8 %, for W and C, respectively. This is very close to the reported $T_c$ in our samples. Also the atomic concentrations are within the range to that of the IBID deposited films. Interestingly, the same $T_c$ of 5 K is observed in RF sputtered W-Si and W-Ge, when the atomic concentrations of either Si or Ge are between 10 % and 30 %.[33] All the above results suggest that superconductivity in W alloys films, regardless of the fabrication method, is attributed to their amorphous structure which is promoted by C, Si, and Ge impurities. This conclusion is further supported by Osofsky *et al*.[37] model which relates this enhancement in $T_c$ to the level of disorder (i.e. resistivity) near the metal-insulator transition of the films.

In conclusion, we have reported on the manifestation of superconductivity at 5.2 K in IBID deposited thin films from Tungsten Carboxyl precursor. The films are amorphous, with atomic concentrations of 40, 40, and 20 %, for W, C, and Ga, respectively. The origin of superconductivity is attributed to the enhanced amorphousity of the Tungsten films by the presence of Carbon and Gallium. By the same mechanism, we expect that other IBID deposited amorphous films, such as Mo and Re, will also show superconducting properties at reasonably high temperatures. We believe that the presented IBID technique of superconducting materials have some advantages over photo/EB lithography for fabricating devices, such as SQUIDS, in the case of inconveniences with using resists, e.g. damage and/or contamination to samples, the need to repair only portions of the circuits, and the ability to deposit superconducting films at different surface orientations on a 3D sample. For example it is possible to simultaneously fabricate two SQUIDS at perpendicular directions. Furthermore, from the progress achieved in the deposition of 3D nanostructures by IBID,[6] this technique can also be applied to the fabrication of 3D superconducting devices and circuits, such as three-dimensional superconducting pick-up coils.

The authors would like to thank K. Nishida and T. Kimura for the EPMA analysis, and T. Aoyagi for his technical help with FIB.

TABLE I. Electron probe microanalysis (EPMA) of an IBID fabricated superconducting thin film. The film dimensions are 200 by 200 μm, in length and width, and 0.12 μm in thickness. The EPMA data are collected at both the center and edge of the film. The average of the two values is also presented. This data show that the concentration of the elements, W, C, and Ga, is essentially uniform over the whole film.

| Atomic ratio (%) | Carbon (C) | Tungsten (W) | Gallium (Ga) |
| --- | --- | --- | --- |
| Center | 40.96 | 39.05 | 19.98 |
| Edge | 39.42 | 39.95 | 20.73 |
| Average | 40.19 | 39.45 | 20.36 |

Table I. E. S. Sadki *et al*.

**Figure captions**

FIG. 1. Scanning electron microscopy picture of a superconducting wire, synthesized by IBID, in a four-point contact configuration for transport measurements. Its dimensions are 10 µm, 300, and 120 nm, in length, width, and thickness, respectively. Scale bar is 2 µm.

FIG. 2. Resistivity versus temperature for magnetic fields applied perpendicular to the superconducting wire from 0 to 5 Tesla.

FIG. 3. Upper critical field versus temperature deduced from the resistivity data. The dashed line is a fit to a power law. The upper critical field, $B_{c2}(0)$, and the coherence length, $\xi(0)$, are 9.5 T and 5.9 nm, respectively.

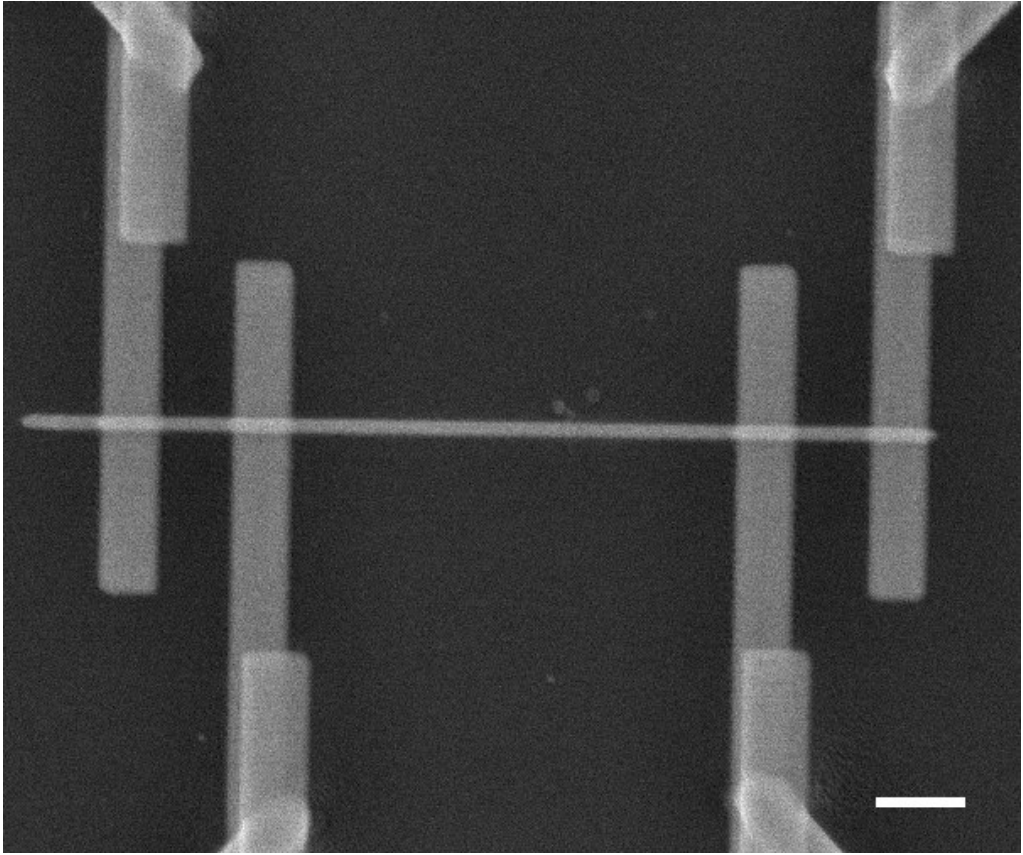

Fig. 1. E. S. Sadki *et al*.

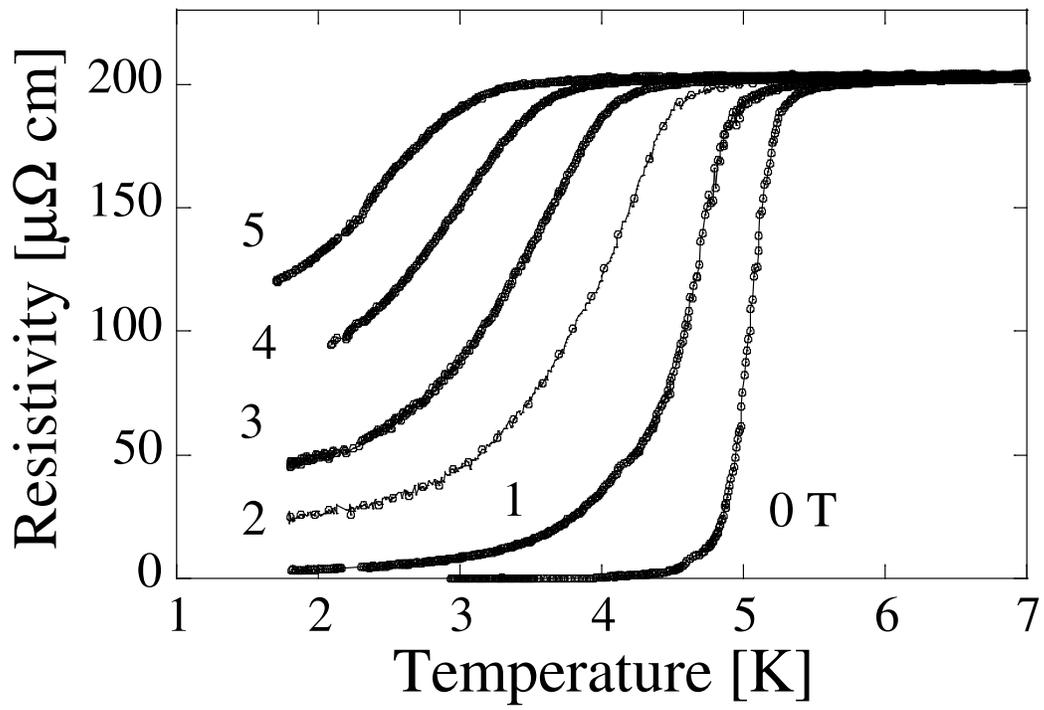

Fig. 2. E. S. Sadki *et al*.

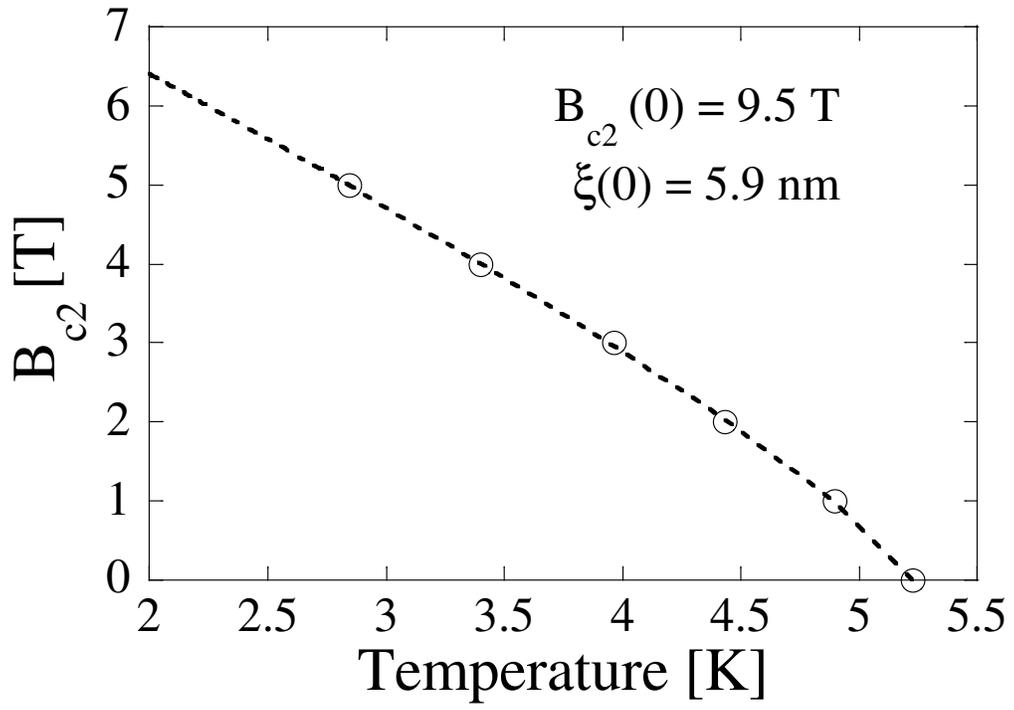

Fig. 3. E. S. Sadki *et al.*